\documentclass[preprint,12pt]{elsarticle}

\usepackage{graphicx}
\usepackage{amssymb}
\usepackage{natbib}

\journal{Advances in Space Research}

\begin{document}

\begin{frontmatter}




\title{Goddard Robotic Telescope\\
- Optical Follow-up of GRBs and Coordinated Observations of AGNs -}


\author[cresst,umbc,nasa]{T. Sakamoto}
\author[fgcu]{C. A. Wallace}
\author[cresst,umcp,nasa]{D. Donato}
\author[nasa]{N. Gehrels}
\author[jhu,nasa]{T. Okajima}
\author[gw,nasa]{T. N. Ukwatta}

\address[cresst]{CRESST and NASA Goddard Space Flight Center, Greenbelt, MD 20771, U.S.A.}
\address[umbc]{Joint Center for Astrophysics, University of Maryand, Baltimore County, Baltimore, MD 21250, U.S.A.}
\address[nasa]{NASA Goddard Space Flight Center, Greenbelt, MD 20771, U.S.A.}
\address[fgcu]{Florida Gulf Coast University, Fort Myers, FL 33965-6565, U.S.A.}
\address[umcp]{Department of Astronomy, University of Maryland, College Park, MD 20742, U.S.A.}
\address[jhu]{Department of Physics and Astronomy, Johns Hopkins University, 3400 North Charles Street, Baltimore, MD 21218, U.S.A.}
\address[gw]{Department of Physics, The George Washington University, Washington, D.C. 20052, U.S.A.}

\begin{abstract}
Since it is not possible to predict when a Gamma-Ray Burst (GRB) will occur 
or when Active Galactic Nucleus (AGN) flaring activity starts, follow-up/monitoring
ground telescopes must be located as uniformly as possible all over
the world in order to collect data simultaneously with {\it Fermi}
and {\it Swift} detections.  However, there is a distinct gap in 
follow-up coverage of telescopes in the eastern U.S. region based on 
the operations of {\it Swift}.  Motivated by this fact, 
we have constructed a 14$^{\prime\prime}$ fully automated optical 
robotic telescope, Goddard Robotic Telescope (GRT), at the Goddard 
Geophysical and Astronomical Observatory.  
The aims of our robotic 
telescope are 1) to follow-up {\it Swift}/{\it Fermi} GRBs 
and 2) to perform the coordinated optical observations of 
{\it Fermi} Large Area Telescope (LAT) AGN.  
Our telescope system consists of off-the-shelf hardware. 
With the focal reducer, we are able to match the field of view 
of {\it Swift} narrow instruments (20$^{\prime} \times 20^{\prime}$).
We started scientific observations 
in mid-November 2008 and GRT has been fully remotely operated since August 2009.  
The 3$\sigma$ upper limit in a 30-second exposure in the R filter is $\sim$15.4 mag; 
however, we can reach to $\sim$18 mag in a 600-second exposures.  Due to the weather 
condition at the telescope site, our observing efficiency 
is 30-40\% on average.  

\end{abstract}


\end{frontmatter}


\section{Introduction}

The $\gamma$-ray emission which comes from extremely energetic $\gamma$-ray 
sources such as Gamma-Ray Bursts (GRBs) and Active Galactic Nuclei (AGNs) 
(especially blazars) is believed 
to be produced by accelerated particles in a relativistic jet with bulk 
Lorentz factors of $\sim$100 for GRBs and $\sim$10 for AGNs.  
A GRB jet is formed when a massive star ($> 10 M_\odot$) collapses into a 
black hole \citep{woodsley1993,pac1998,mac1999}.  On the other hand, an AGN jet is a continuous outflow 
from an active supermassive black hole ($10^{6-9}$ $M_\odot$) \citep[e.g.][]{urry1995}.
Although the energy scale and bulk motion of jets differ by an 
order of magnitude between GRBs and AGNs, their radiation processes
are expected to be similar.  Therefore, understanding radiation processes
in the context of shock physics and particle acceleration using both
GRBs and AGNs will provide a deeper understanding of the
the fundamental physics in these extreme environments.

Recent observations of prompt GRB optical emission by 
ground/space robotic telescopes are providing key data 
for understanding the radiation mechanism of GRBs. The observations 
of GRB 050820A by RAPid Telescope for Optical Response
(RAPTOR; \citep{ves2006}) and GRB 061121 by {\it Swift} UV/Optical Telescope (UVOT; 
\citep{page2007})
indicate that there are at least two components in
the prompt optical emission.   
One component is optical emission correlated with the prompt $\gamma$-ray
emission.  The other component is a smoothly rising and decaying component 
during the prompt $\gamma$-ray phase.  The first component could be 
interpreted as emission from an internal shock because of the similar 
variability between the optical and $\gamma$-ray bands.  The second component
could be due to an external shock interacting with the inter-stellar
matter. However, the Robotic Optical Transient Search Experiment (ROTSE) 
has shown that there are a couple of cases where the 
early optical emission does not correlate with the $\gamma$-ray emission 
(e.g. \citep{rykoff2005}).
The extremely bright prompt optical 
emission from GRB 080319B observed by TORTORA and 'PI of the sky' challenges 
the standard picture of the GRB emission model \citep{racusin2008}.  

Blazars form a sub-group of radio-loud AGNs and show an extreme variability
at all wavelengths \citep{urry1999}.  The most accepted scenario is that a rotating
supermasive black hole surrounded by an accretion disk with an intense
plasma jet closely aligned to the line of sight is responsible for the
blazar emission.  However, fundamental understanding of the radiation
process in blazars requires extensive monitoring
campaigns at all wavelengths.  In particular, properties of variability
(including flares) and spectra from simultaneous data in various
wavelengths provides key information for a physical understanding of blazars.

\section{Motivation and Telescope Specification}

\begin{figure}[t]
\begin{center}
\includegraphics[height=10cm]{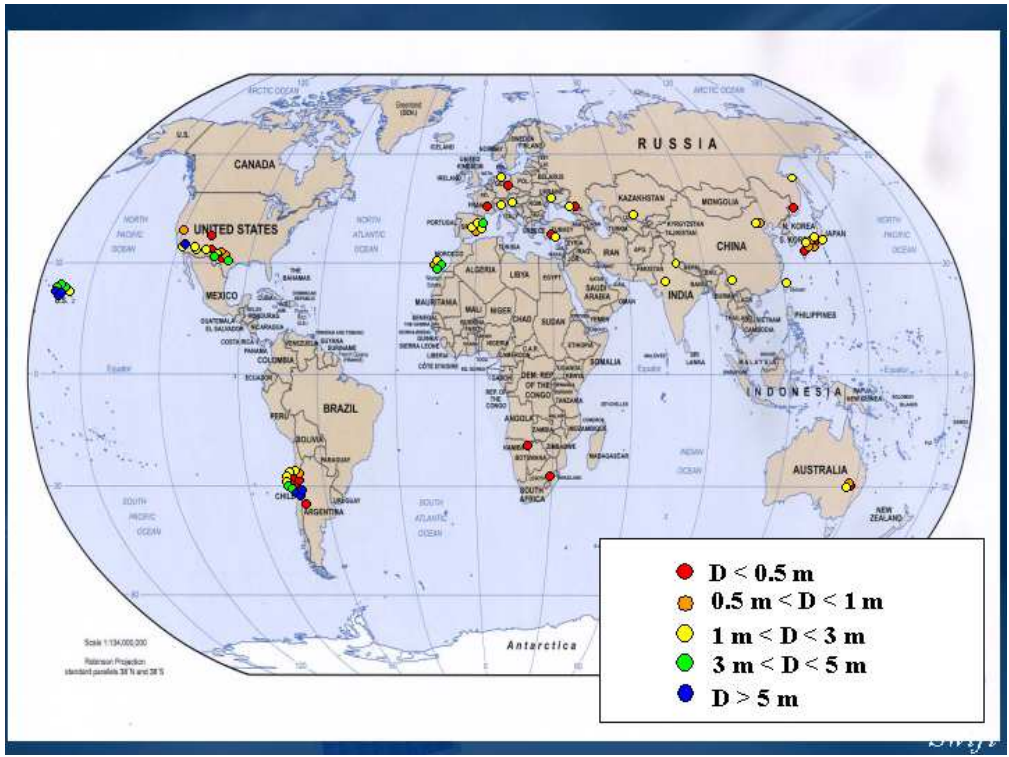}
\caption{\label{tel_map} Distribution of the 
GRB follow-up telescopes (information from GCN Circular of the Swift GRBs 
from September 2006 to March 2007).}
\end{center}
\end{figure}

Since it is not possible to predict when a GRB will occur or when 
AGN flaring activity starts, follow-up/monitoring
ground telescopes must be located as uniformly as possible all over
the world in order to collect data simultaneously with {\it Fermi}
and {\it Swift} detections.  Based on the operations of 
{\it Swift}, however, we notice a distinct gap in follow-up coverage
in the eastern U.S. region (Figure \ref{tel_map}).  
This fact motivated us to construct a fully automated optical telescope
at the Goddard Space Flight Center.  If there is no GRB to observe, 
this telescope will perform extensive optical monitoring of {\it Fermi} 
Large Area Telescope (LAT) AGNs to identify the variability on the time 
scale of a day \citep{totani2005}.

\begin{figure}
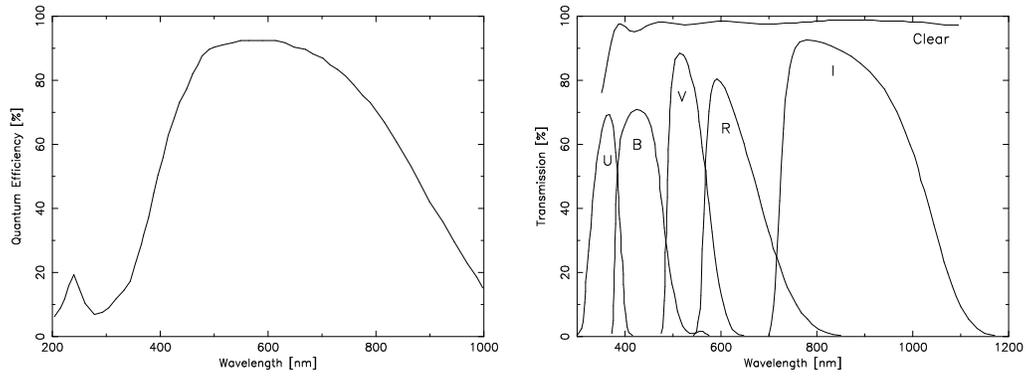

\begin{center}
\includegraphics[height=6.5cm,angle=-90]{U47_QE.ps}
\hspace{0.2cm}
\includegraphics[height=6.5cm,angle=-90]{trans_all.ps}
\caption{\label{qe_trans} Left: The quantum efficiency of U47 CCD camera.  The CCD chip is CCD47-10 
(back illuminated CCD) by e2V technologies.  Right: The transmission curves of each filter.}
\end{center}
\end{figure}

\begin{figure}
\begin{minipage}{7cm}
\begin{center}
\includegraphics[height=7.0cm,clip]{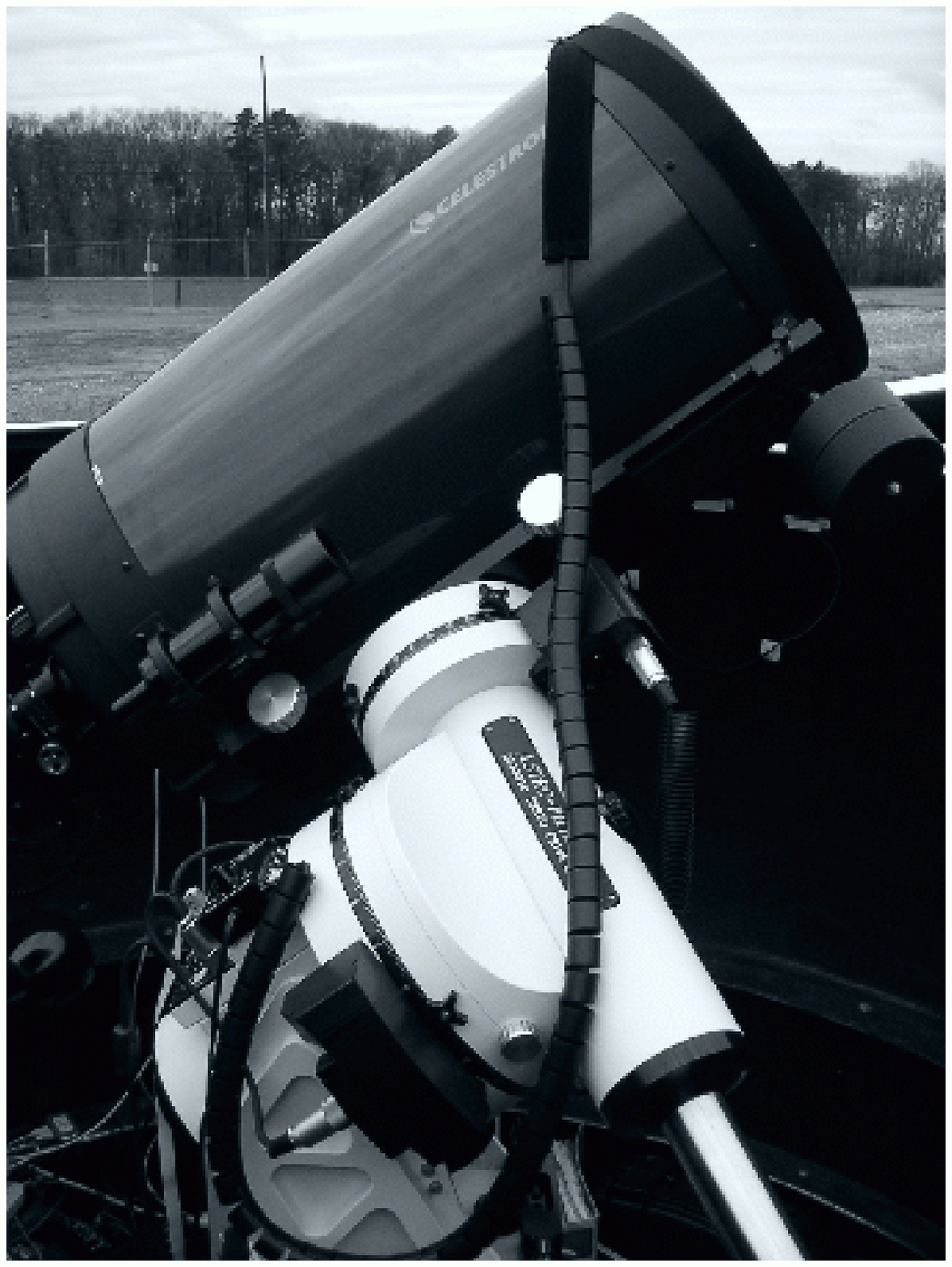}
\caption[short caption for figure 2]{\label{tel} Picture of the telescope.}
\end{center}
\end{minipage}
\hspace{0.5cm}
\begin{minipage}{7cm}
\begin{center}
\includegraphics[height=7cm,clip]{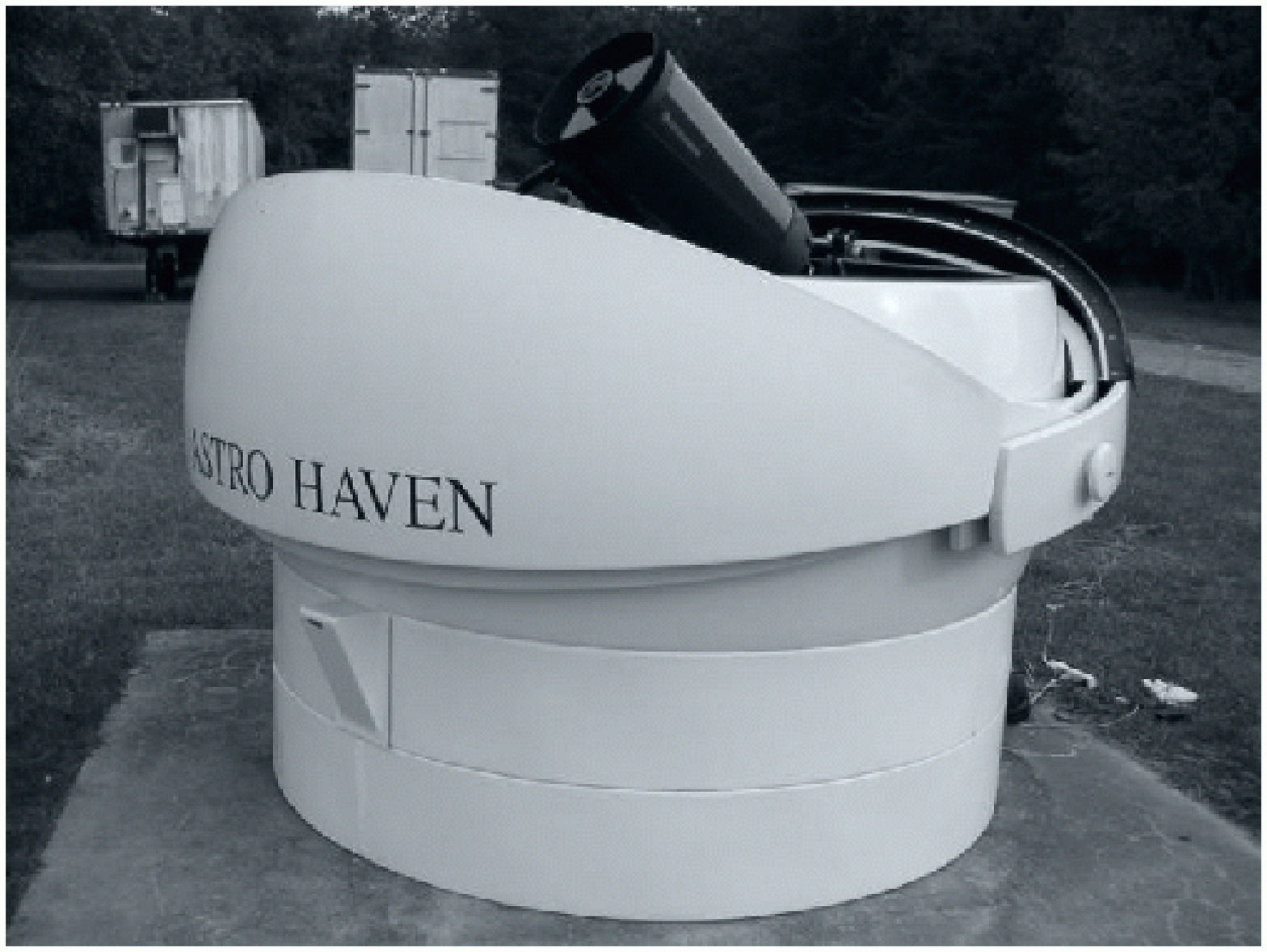}
\caption[short caption for figure 3]{\label{dome} Picture of the dome with the telescope inside.}
\end{center}
\end{minipage}
\end{figure}

The Goddard Robotic Telescope (GRT) system consists of a 
14$^{\prime\prime}$ Celestron Optical Telescope Assembly (OTA), 
an Astro-Physics 
1200GTO mount, an Apogee U47 CCD camera, a JMI electronic focuser 
(EV2CM) with a PC controller (PCFC), and a Finger Lake Instrumentation 
color filter wheel (CFW-1-8) with standard Johnson U, B, V, R, I and 
Clear filters.  The quantum efficiency of the camera and the transmission 
curves of each filter are shown in Figure \ref{qe_trans}.  
With the focal 
reducer (Celestron f/6.3), a 20$^{\prime} \times 20^{\prime}$ field of view 
has been achieved.  The observatory dome is an Astro Haven 7ft clam-shell dome.  
Opening and closing of the dome is controlled by the serial interface.  
Figures \ref{tel} and \ref{dome} are pictures of the telescope and the dome.  

The dome, the weather station (Davis Instruments Vantage Pro2), the webcams 
(Logitech Quick Cam Chat), and the rain sensors are controlled by another 
Linux PC.  The dome is directly connected by a serial port to the PC.  
The Linux-based weather station control software (wview\footnote{http://www.wviewweather.com/}) 
and the 
webcam driver (spca5\footnote{http://mxhaard.free.fr/index.html}) are used to control the weather station and 
the webcams.  The signal from rain sensors (Asuzac Inc.) is recorded by the I/O 
board (Arduino\footnote{http://www.arduino.cc/}), and used to close the dome automatically in case 
of rain.  

The telescope site is the Goddard 
Geophysical and Astronomical Observatory 
(GGAO) which is about 1.5 miles northeast from the Goddard main campus.  

\section{Automation of the Observatory}

The whole telescope system, including the telescope mount, the focuser, 
the CCD camera and the filter wheel, is controlled by the ACP Observatory 
Control Software.\footnote{http://acp.dc3.com/index2.html}  An advantageous feature of the ACP software 
is to perform a short test exposure (five seconds in our case) after each slew and 
to do astrometry automatically 
using the Guide Star Catalog\footnote{A user can specify either the Guide Star 
Catalog 1.1 or the USNO A2.0 catalog.} to find the center 
sky coordinate of an image.  
If the pointing is off, ACP requests the mount to re-slew so that 
the requested position will be exactly at the center of an image.  Furthermore, ACP 
will automatically construct a pointing model 
by recording an offset between the requested position and the actual slew 
position.  This pointing model will be used for every slew to achieve a better 
pointing at the initial attempt.  

We developed a script to make an observation schedule for a single night.  
This scheduler calculates the elevation of all objects in the database.  
It selects objects which are visible from the site (an elevation has to be 
larger than 35$^{\circ}$ to avoid the light pollution), 
and then schedules objects at the time when they reach 
the highest elevation.  The output of the script is a text file with lines 
of ACP commands.  ACP commands consist of an object name, 
a set of coordinates for each object, filters, exposure times of each filter, a bin size of 
images and the starting time of the observation.  
Since it is a text file, it is easy to modify the schedule by hand if we need to 
schedule an object which is not in the database.  The sky flat, bias and dark 
frames are scheduled at the end of observations.  The script runs once a day 
in the afternoon.  
This schedule file is automatically 
submitted through ACP by the observation control script which we describe next.   
As of September 2010, we have 58 {\it Fermi} LAT AGNs, 70 variable stars, 33 near-by 
galaxies to search for supernova, and 37 guest observers' objects in the database.  

A script to control the whole observatory has been developed.  
The script calculates the Sun's elevation every 10 minutes and sends 
commands to ACP or to the dome at specific Sun elevations to perform 
actions such as opening/closing the dome, changing the CCD 
temperature, submitting the schedule or re-submitting the schedule in case 
of a termination of on-going observation for various reasons.  

The GRB position information is received via a Gamma-ray bursts Coordinates 
Network (GCN) socket connection.  
Once the GRB position has been received, our custom GCN receiver code will 
calculate the elevation of the GRB at the site, and if it is observable 
(an elevation of a GRB has to be higher than 20$^{\circ}$ and also the Sun's 
elevation has to be less than $-20^{\circ}$), 
it will send commands to ACP to quit on-going observations 
and execute a GRB observation.  
We collect 200 frames of 5 sec, 80 frames of 30 sec, 120 frames of 
60 sec and 72 frames of 100 sec exposure of the GRB field in the time order 
given.  
Based on a real-time observation of several GRBs, GRT took 47 sec 
on-average to terminate on-going observations and complete a slew to the 
GRB location.  The initial exposure was made 66 sec on average after 
we received the GRB position information through the GCN.  

\begin{figure}[h]
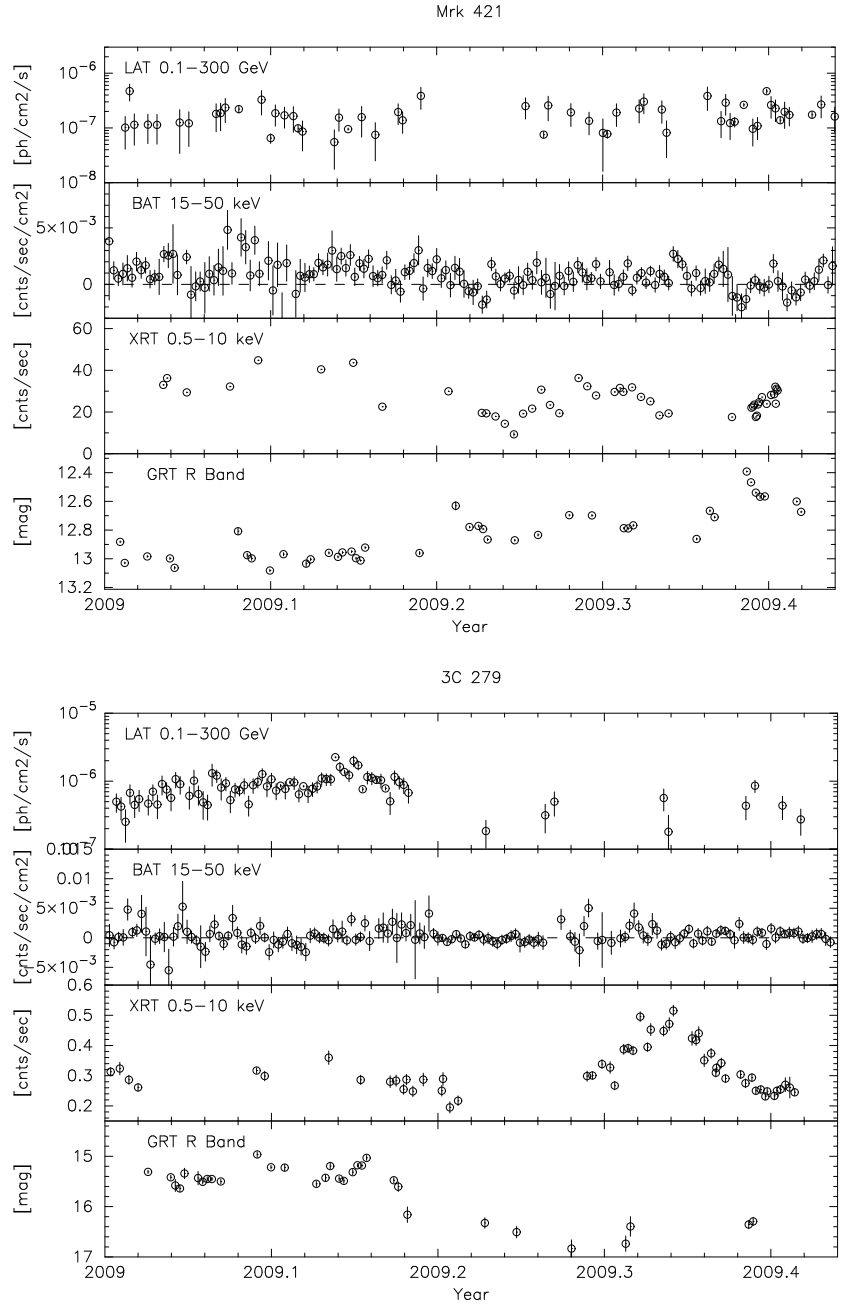

\begin{center}
\includegraphics[height=11cm,angle=-90]{Mrk_421_multi_lc_mod2.ps}
\end{center}
\begin{center}
\includegraphics[height=11cm,angle=-90]{3C279_multi_lc_mod2.ps}
\end{center}
\caption{Multi-wavelength light curves of Mrk 421 (top) and 3C 279 (bottom).  From top to 
bottom for each source, the light curve in the 0.1--300 GeV band by {\it Fermi}/LAT, 
the 15--50 keV band by {\it Swift}/BAT, the 0.5--10 keV band by 
{\it Swift}/XRT, and the optical R band by GRT.\label{grt_lc}}
\end{figure}

The automatic script for LAT AGNs has also been 
developed to do the photometry. The script performs the relative photometry using 
Sextractor\footnote{http://terapix.iap.fr/rubrique.php?id\_rubrique=91/} 
and PyRAF\footnote{http://www.stsci.edu/resources/software\_hardware/pyraf} 
adopting three reference stars defined 
for each object.  The aperture photometry (IRAF phot task) is obtained using 
a circular extraction region of 5$^{\prime\prime}$ for the source and an annulus with 
inner/outer radius of 5.5$^{\prime\prime}$/6.25$^{\prime\prime}$ for the background.
The script also extracts the publically available light curves of {\it Fermi} 
LAT,\footnote{http://fermi.gsfc.nasa.gov/ssc/data/access/lat/msl\_lc/} 
{\it Swift} BAT,\footnote{http://swift.gsfc.nasa.gov/docs/swift/results/transients/} 
and {\it Swift} XRT\footnote{http://www.swift.psu.edu/monitoring/} 
of the AGNs.   
Examples of multi-band AGN light curves produced by this script are shown in 
Figure \ref{grt_lc}.  All of the AGN light curves observed by GRT are available 
to the public from the dedicated web page.\footnote{http://asd.gsfc.nasa.gov/Takanori.Sakamoto/GRT/index.html} 

\begin{figure}[h]
\begin{center}
\includegraphics[height=7cm]{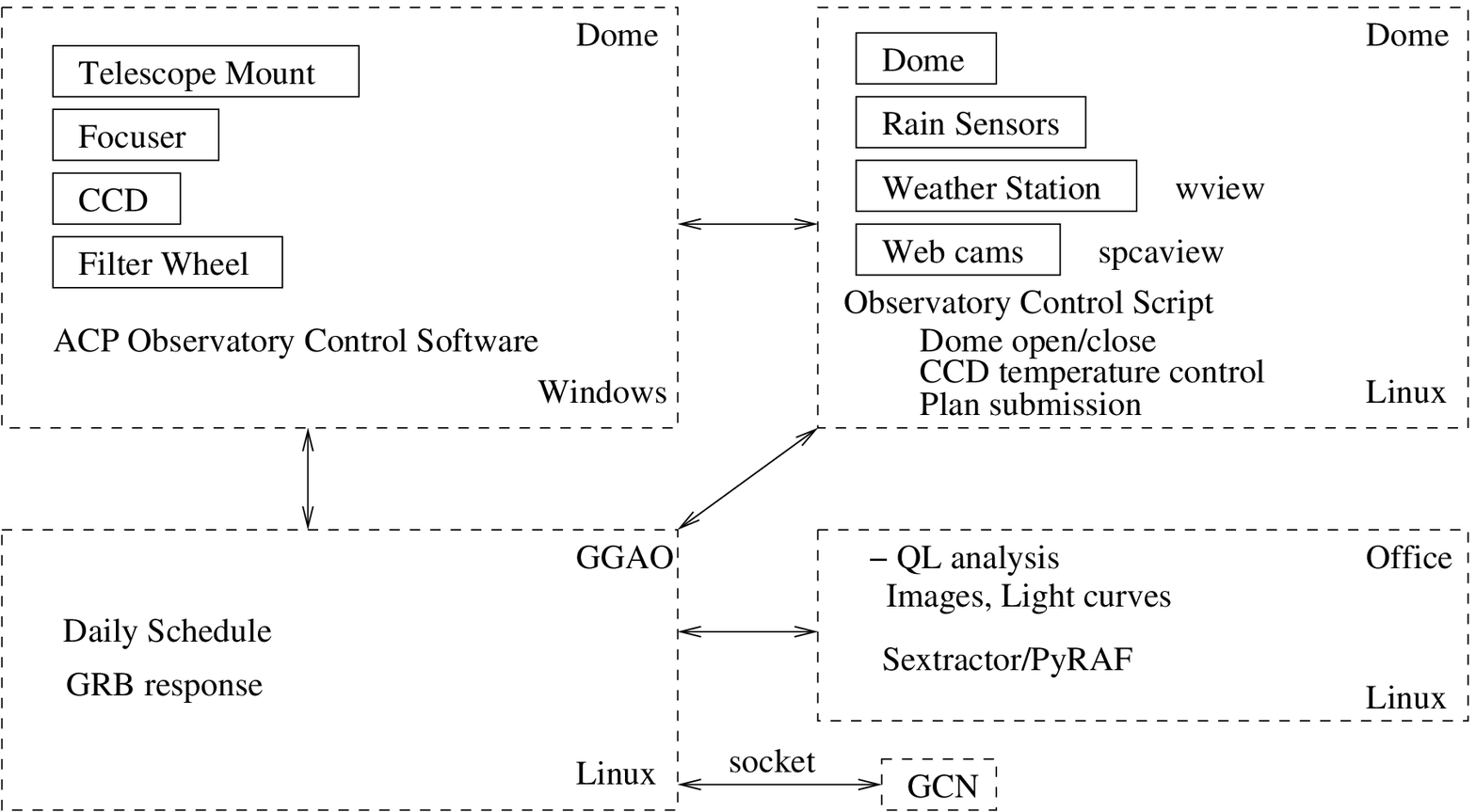}
\caption{\label{grt_flowchart} Flowchart of GRT automation.}
\end{center}
\end{figure}

In Figure \ref{grt_flowchart}, we summarize the automation of the observatory 
as a flowchart.  

\section{Performance}

\begin{figure}
\begin{center}
\includegraphics[height=11cm,angle=-90]{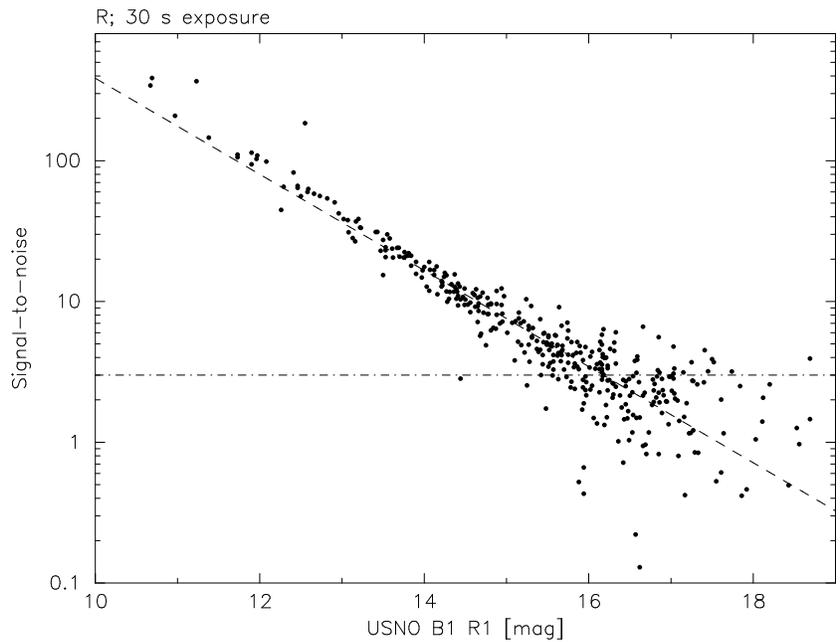}
\caption{\label{detlimit_30s} Relationship between signal-to-noise ratio and the R1 magnitude 
of the USNO B1 catalog for the image of the field of GRB 100915 at a 30 s exposure in the R filter.  
The horizontal dash-dotted line shows the 3$\sigma$ detection limit.}
\end{center}
\end{figure}

\begin{figure}
\begin{center}
\includegraphics[height=11cm,angle=-90]{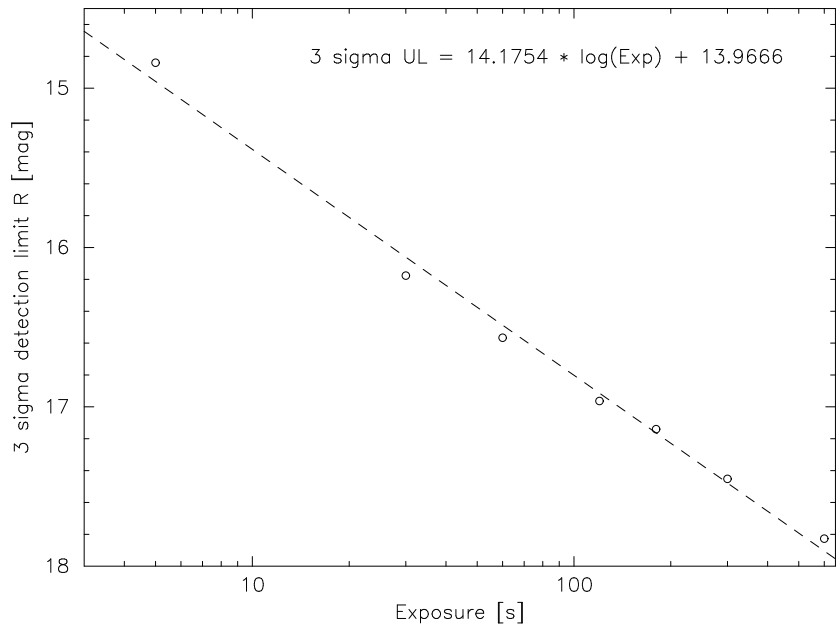}
\caption{\label{detlimit_exp} Relationship between 3$\sigma$ detection limit and exposure time.  
The best fit relation (dashed line) is 3$\sigma$ upper limit = 14.1754 $\times$ $\log$(exposure 
time)+13.9666.}
\end{center}
\end{figure}

We used the observations of GRB 100915 to estimate the detection limit 
of our telescope system.  The first several hours 
had good weather conditions.  The CCD temperature was $-20^{\circ}$C.  
The observations 
were performed in the airmass range from 1.13 to 1.27 (about four hours) 
collecting 423 frames in total with four different exposures (5 sec, 30 sec, 
60 sec and 100 sec).  All the light frames are in the R filter and corrected for the bias, 
the dark and the flat.  First, we extract the 
sources in the image using Sextractor.  Then, we perform photometry 
using the IRAF phot task.  
The aperture foreground radius is 1.6 times the averaged FWHM of extracted sources.  
The background annulus with inner and outer radius are, respectively, 1.75 and 2.0 
times the averaged FWHM of the extracted sources.  
A signal-to-noise ratio of extracted sources 
is calculated using the IRAF phot result.\footnote{The signal and the noise are 
calculated by 
\[signal = SUM - AREA \times MSKY,\]
and 
\[noise = \sqrt{\frac{SUM-AREA \times MSKY}{GAIN} + 1.23 AREA \times STDEV^{2}}.\]
}  The USNO B1 catalog R1 magnitude is used as the R magnitude of the extracted 
sources.  The 3$\sigma$ detection limit has been defined by fitting the signal-to-noise 
versus R magnitude data, and finding the 3$\sigma$ crossing point to the fit (see Figure 
\ref{detlimit_30s}).  The 3$\sigma$ detection limit of a 30 sec exposure in R is 
$\sim$15.4 mag.  To understand the detection limit as a function of exposure time, 
we repeated the same analysis of the images with different exposures (5 sec, 30 sec, 
60 sec, 120 sec, 180 sec, 300 sec and 600 sec).  We stacked 30 sec images using 
swarp\footnote{http://terapix.iap.fr/rubrique.php?id\_rubrique=49} to create images 
with $>$30 sec exposure.  The relationship between 
the 3$\sigma$ detection limit and exposure time is shown in Figure \ref{detlimit_exp}.  
We reached $\sim$18 mag in a 600 sec exposure.  

\begin{figure}
\begin{center}
\includegraphics[height=6.7cm,angle=-90]{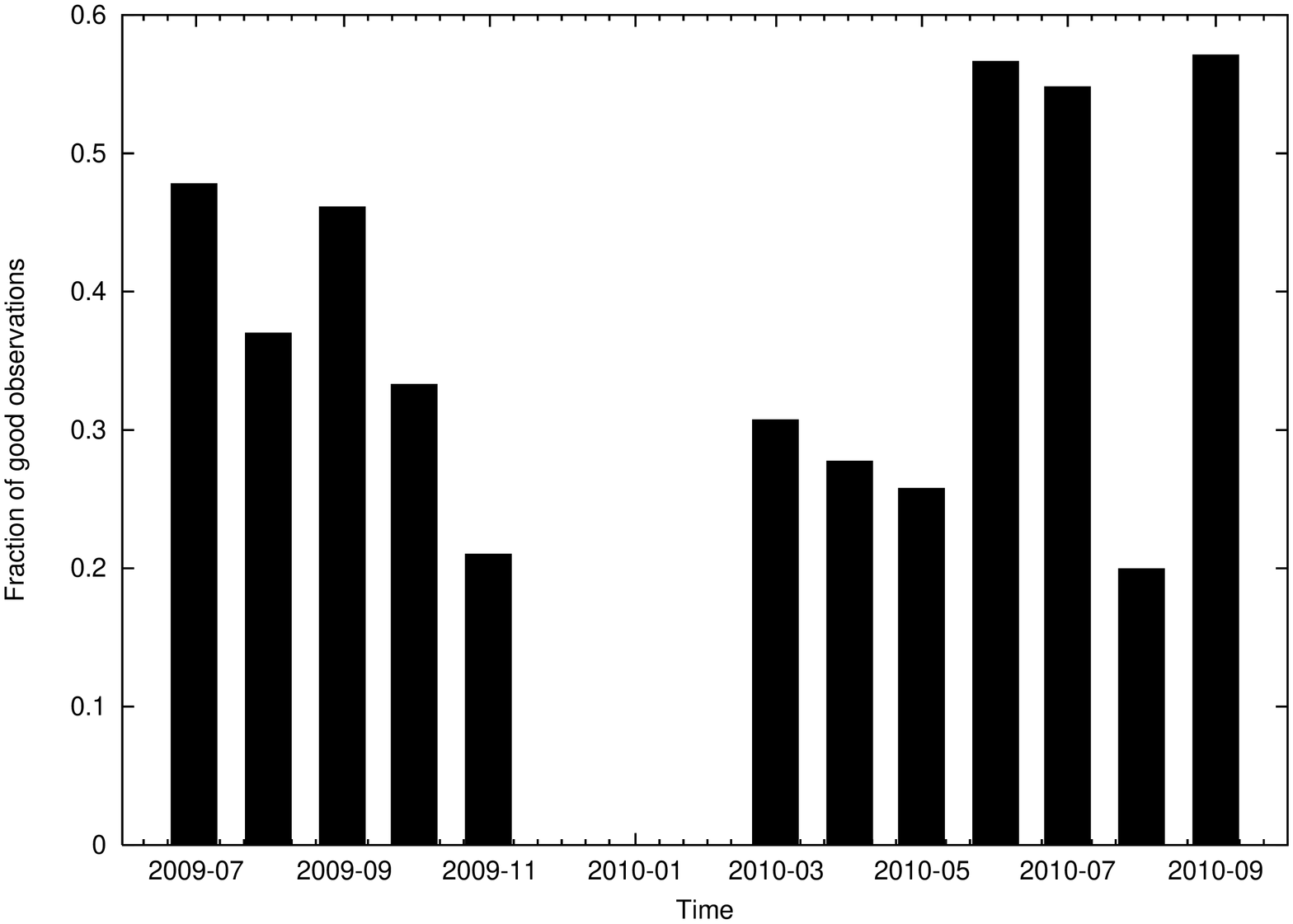}
\includegraphics[height=6.7cm,angle=-90]{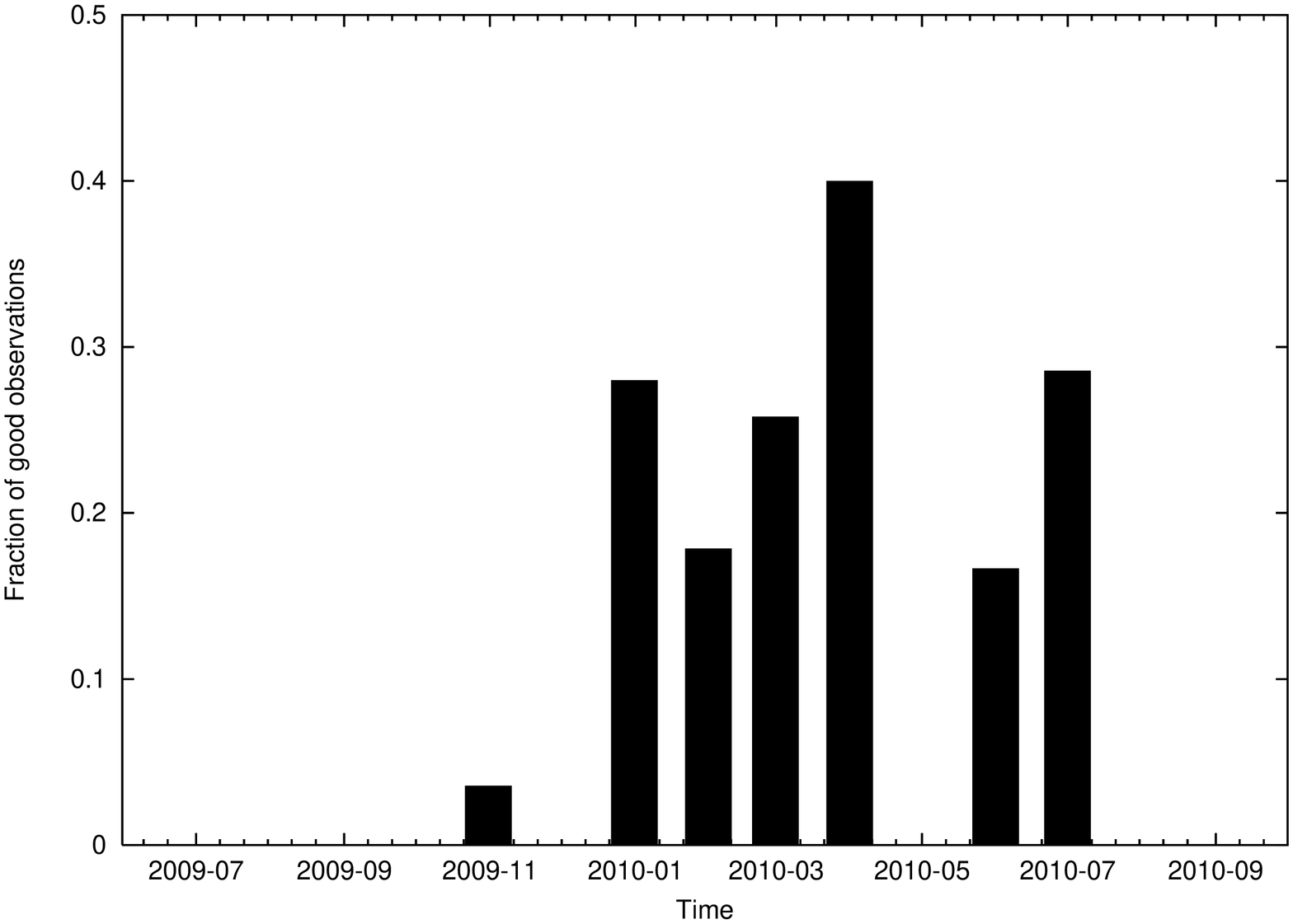}
\end{center}
\begin{center}
\includegraphics[height=6.7cm,angle=-90]{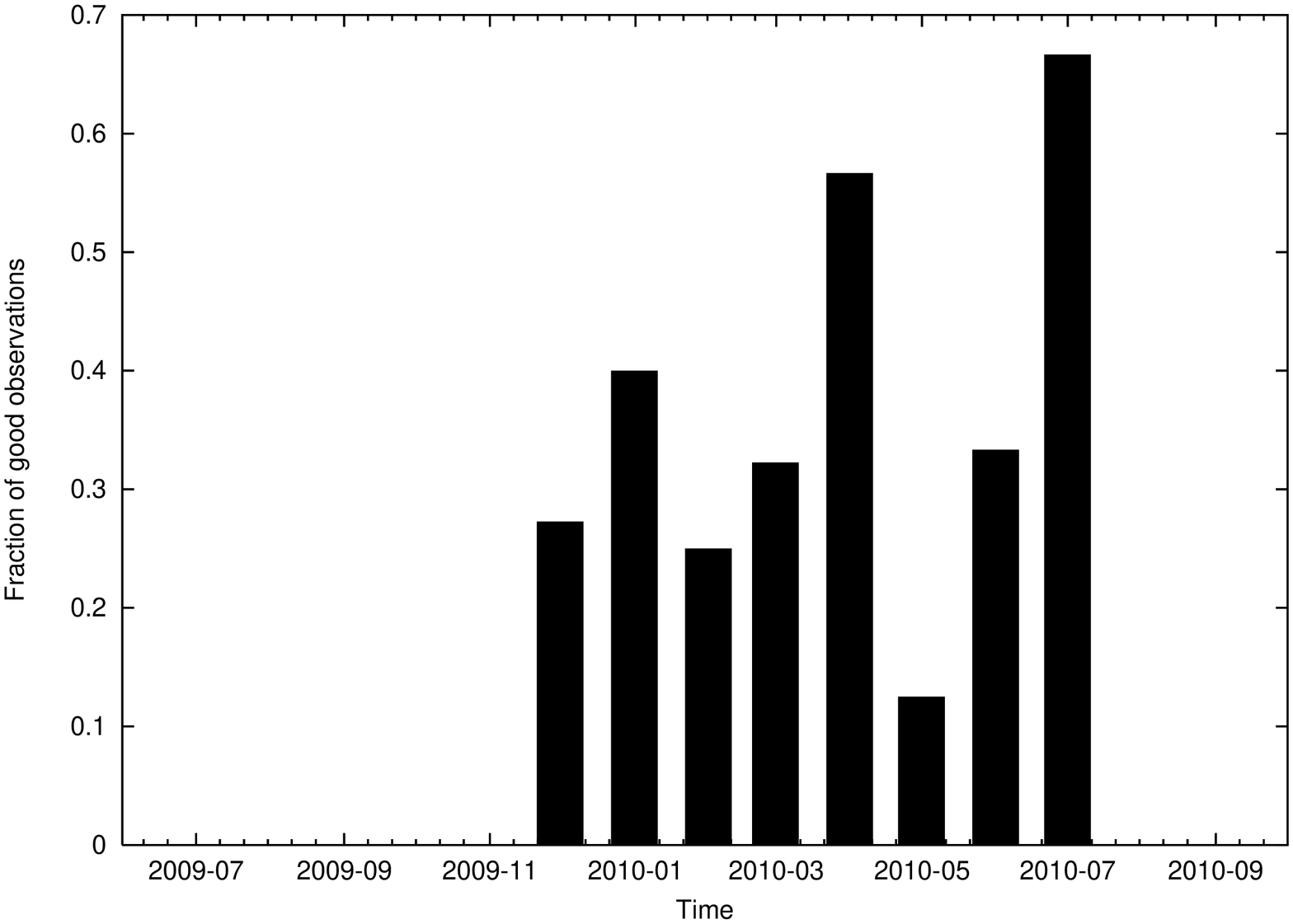}
\includegraphics[height=6.7cm,angle=-90]{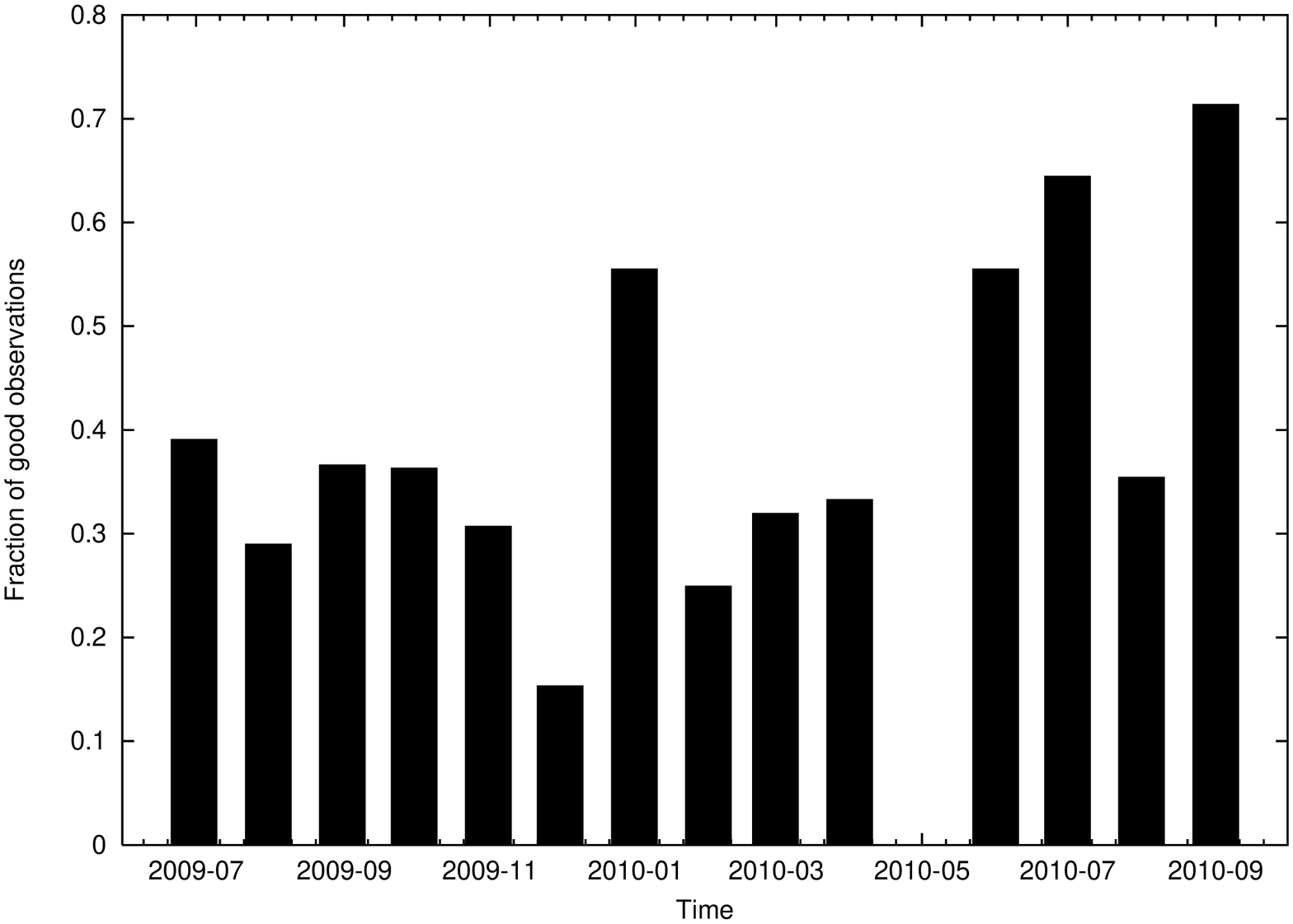}
\end{center}
\caption{\label{obs_good_frac} Fraction of good observations for each month for four AGN 
observations: Mrk 501 (top left), Mrk 421 (top right), 3C273 (bottom left) 
and 1ES 1959+650 (bottom right).}
\end{figure}

In Figure \ref{obs_good_frac}, we summarize the fraction of good observations 
for each month for four AGN observations.  Although we experienced few months 
with $>$60\% of good weather night, the typical fraction of observable nights 
is 30-40\%.

\section{Current Status}

We started fully automatic operation in mid-November 2008.  
Although, at the beginning, we had conservative operations by having 
a duty operator every night, GRT is completely 
human free as of August 2009.  We are currently observing $\sim$85 objects and collecting 
$\sim$350 images per night.  

Our optical contribution to the multi-wavelength observing campaign of 
the $\gamma$-ray outburst of 3C 454.3 has been published \citep{agile_3c454}.  Further 
collaborations  
with the {\it Fermi}/LAT team for Mrk 421, Mrk 501, 1ES 1959+650 and MG4 J200112+4352 are 
on-going.  

We have performed 13 follow-up observations of GRBs (12 {\it Swift} GRBs 
and one {\it INTEGRAL} GRB).  Three of them were observed within 10 minutes after 
the GRB trigger.  Although there is no detection of prompt optical 
emission or an afterglow so far, 
we confirmed that GRT can achieve a 17.5 to 19 magnitude limit in the R filter, and 
also respond automatically and promptly to a GRB through the GCN socket connection.  

\section{Acknowledgements}

We would like to thank the anonymous reviewers for comments and suggestions that 
materially improved the paper.  We also would like to thank H.~A. Krimm for carefully 
reading the paper.  












\end{document}